\documentclass[prl,twocolumn,showpacs,superscriptaddress, floatfix, reprint]{revtex4-1}

\usepackage{graphicx}
\usepackage{amssymb}
\usepackage{amsmath}
\usepackage{amsfonts}
\usepackage{bm}
\usepackage{color}
\usepackage{hyperref}

\newcommand{\op}[1]{\hat{#1}}

\newcommand{\be}{\begin{equation}}
\newcommand{\ee}{\end{equation}}

\begin{document}

\title{An impurity immersed in a double Fermi Sea}

\author{R. Alhyder}%
\email{Corresponding author: ragheed.alhyder@lkb.ens.fr}
\affiliation{Laboratoire Kastler Brossel, ENS-Universit\'e PSL, CNRS, Sorbonne Universit\'e, Coll\'ege de France.%\\This line break forced% with \\
}%
\affiliation{Laboratoire de physique de l'Ecole Normale sup\'erieure, ENS, Universit\'e PSL, CNRS, Sorbonne Universit\'e, Universit\'e de Paris, F-75005 Paris, France.%\\This line break forced% with \\
}%

\author{X. Leyronas}%
\affiliation{Laboratoire de physique de l'Ecole Normale sup\'erieure, ENS, Universit\'e PSL, CNRS, Sorbonne Universit\'e, Universit\'e de Paris, F-75005 Paris, France.%\\This line break forced% with \\
}%

\author{F. Chevy}%
\affiliation{Laboratoire Kastler Brossel, ENS-Universit\'e PSL, CNRS, Sorbonne Universit\'e, Coll\'ege de France.%\\This line break forced% with \\
}%

\begin{abstract}
We present a variational calculation of the energy of an impurity immersed a double Fermi sea of non-interacting Fermions. We show that in the strong-coupling regime, the system undergoes a first order transition between  polaronic and trimer states. Our result suggests that the smooth crossover predicted in previous literature for a superfluid background is the consequence of Cooper pairing and is absent in a normal system.
\end{abstract}

\maketitle

%\section{Introduction}
Introduced for the first time by Landau and Pekar to describe transport properties of electrons in semi-conductors \cite{landau1948effective},  polaron physics has become a prototype for other impurity problems in quantum many-body systems, from solid state \cite{kondo1964resistance,anderson1961localized} to nuclear physics \cite{zuo20041s0}. More recently, the crucial role played by polaronic properties in the performances of solar panels has revived the interest for this system in  semi-conductors, which is now an active field of research in both applied mathematics \cite{seiringer15polaron} and fundamental physics \cite{alexandrov2008polarons}.

% -&\frac{1}{\Omega}\sum_{k'}\frac{D(k')}{\lambda^2+k^2+{k'}^2+\bm k\cdot\bm k'}=0\nonumber

Thanks to the versatility of their experimental investigation tools, ultracold atoms have become over the past decade a remarkable playground for the exploration of quantum many-body physics \cite{bloch2008many,zwerger2012BCSBEC,Chevy2010Unitary,massignan2014polarons}. In this context, polaron physics has been the subject of extensive research, starting from the so called "Fermi polaron", corresponding  to an impurity immersed in a spin polarized Fermi sea \cite{chevy2006upa,lobo2006nsp,prokof'ev08fpb,nascimbene2009pol,schirotzek2009ofp,yan2019boiling} (see also \cite{sidler2017fermi} for its realization in exciton-polariton systems), to the Bose polaron, where the impurity interacts with a weakly interacting Bose-Einstein condensate \cite{jorgensen2016observation,hu2016bose,levinsen2015impurity}. The study of dual superfluids of bosons and fermions recently paved the way to the study of a novel type of polaronic system where the impurity is immersed in a spin 1/2 fermionic superfluid \cite{Ferrier2014Mixture,roy2017two,yao2016observation}. This superfluid version of the Fermi polaron interpolates between aforementioned Fermi and Bose-polarons, and like in this latter  case, three-body physics, and most notably the existence of Efimov trimer states \cite{efimov1970energy,naidon2017efimov} play an important role in shaping the phase diagram of the system \cite{nishida2015polaronic,yi2015polarons,Pierce19few}. Using a mean-field description of the superfluid, the generalization of the Fermi-polaron wavefunction suggested the existence of a crossover between the polaron and trimeron states. In particular, ref. \cite{yi2015polarons} proposed a variational ansatz
\be
|\psi\rangle=\left(\alpha\widehat b_0^\dagger+\sum_{k,\bm k'}\beta_{\bm k,\bm k'}\widehat b_{\bm k'}^\dagger\widehat\gamma_{\bm k,\uparrow}^\dagger\widehat\gamma_{-\bm k'-\bm k,\downarrow}^\dagger\right)|{\rm BCS}\rangle,
\label{Eq:BCSAnsatz}
\ee
where $|{\rm BCS}\rangle$ is the BCS mean-field ground state, $\widehat b_{\bm k} $ is the annihilation operator of an impurity of momentum $\bm k$ and  $\gamma_{\bm k,s}$ that of the Bogoliubov modes of the underlying superfluid. Under this assumption, the crossover arises from the fact that the $\gamma$'s
are linear combinations of creation and annihilation operators of real fermions $\widehat a_{\bm k,s=\uparrow,\downarrow}$. Indeed, the variational state (\ref{Eq:BCSAnsatz}) contains terms proportional to $\widehat b^\dagger\widehat a_{s}^\dagger\widehat a_{-s}^\dagger$ and $\widehat b^\dagger\widehat a_{\bm k,s}^\dagger\widehat a_{\bm k',s}$ that describe respectively a trimer made of fermions above the Fermi surface and a polaron dressed by a particle-hole pair. The crossover is here a direct consequence of the mixing between particles and holes induced by the quantum coherence of the superfluid state and trimers can be interpreted as bound states between the impurity and preexisting Cooper pairs. The question that naturally arises is then the existence of such a crossover in a normal system.   Here, we analyze this question by considering the interaction of an impurity with  an ideal gas of spin 1/2 fermions. By considering  a variational ansatz  incorporating two particle-hole excitations  we suggest that the crossover is suppressed in the absence of Cooper-pairing and is replaced by a sharp (first-order like) transition between a polaron and a trimer branch. We show that this transition is driven by the onset of momentum correlations between holes of the background Fermi seas which can be  associated with two subspaces that are uncoupled by the many-body Hamiltonian and leading to the suppression of the center of mass of the trimer.

%\section{The Model}

We consider an impurity of mass $m_i$ coupled to a Fermi sea of {\em non-interacting} spin 1/2 fermions of mass $m$. We describe the system using a two-channel model known to give rise to Efimov trimers without requiring any additional physical ingredient \cite{gogolin2008analytical}. Assuming periodic boundary conditions in a box of quantization volume $\Omega$, the Hamiltonian then takes the general form

\be
\begin{split}
\op H=&\sum_{\bm k,s} \frac{\hbar^2 k^2}{2m}\op a_{\bm k,s}^\dagger\op a_{\bm k,s}^{\phantom{\dagger}}+\sum_{\bm k} \frac{\hbar^2 k^2}{2m_i}\op b_{\bm k}^\dagger\op b_{\bm k}^{\phantom{\dagger}}+\\&
\sum_{\bm k} \left(\frac{\hbar^2 k^2}{2M}+E_0\right)\op c_{\bm k,s}^\dagger\op c_{\bm k,s}^{\phantom{\dagger}}+\\&
\frac{\Lambda}{\sqrt{\Omega}}\sum_{\bm k,\bm k'}\left[ \op a_{\bm ks}^{\phantom{\dagger}}\op b_{\bm k'}^{\phantom{\dagger}}\op c_{\bm k+\bm k',s}^\dagger+{\rm h.c.}\right],
\end{split}
\ee
where $\op a_{\bm k,s}$ is the annihilation operator of a fermion of spin $s$ and momentum $\hbar\bm k$, $\op b_{\bm k}$ is the annihilation operator of an impurity and $\widehat c_{\bm k,s}$ is the annihilation operator of a molecule made of an impurity and a spin $s$ atom. $E_0$ and $M=m+m_i$ are the binding energy and the mass of the bare molecules. We assume here that the interaction between the impurity and each spin component is the same. The coupling $\Lambda$ does not depend on momentum, but a UV-cutoff $k_c$ is introduced to match the scattering length $a$ and the effective range $R_e$ of the true potential using the following relations \cite{gogolin2008analytical}
% Modified by Ragheed
% \be
% \frac{1}{a}=\frac{2k_c}{\pi}+\frac{2\pi\hbar^2E_0}{m^*\Lambda^2}\hskip 1cm
% R_e=\frac{8\pi\hbar^4}{{m^*}^2\Lambda^2},
% \ee
\be
\frac{1}{a}=\frac{2k_c}{\pi}-\frac{2\pi\hbar^2E_0}{m^*\Lambda^2}\hskip 1cm
%R_e=\frac{2\pi\hbar^4}{{m^*}^2\Lambda^2},
R_e=\frac{\pi\hbar^4}{{m^*}^2\Lambda^2},
\ee
where $m^*$ is the reduced mass of the impurity/fermion pair.

\begin{figure}
    \centering
    \includegraphics[width=\columnwidth]{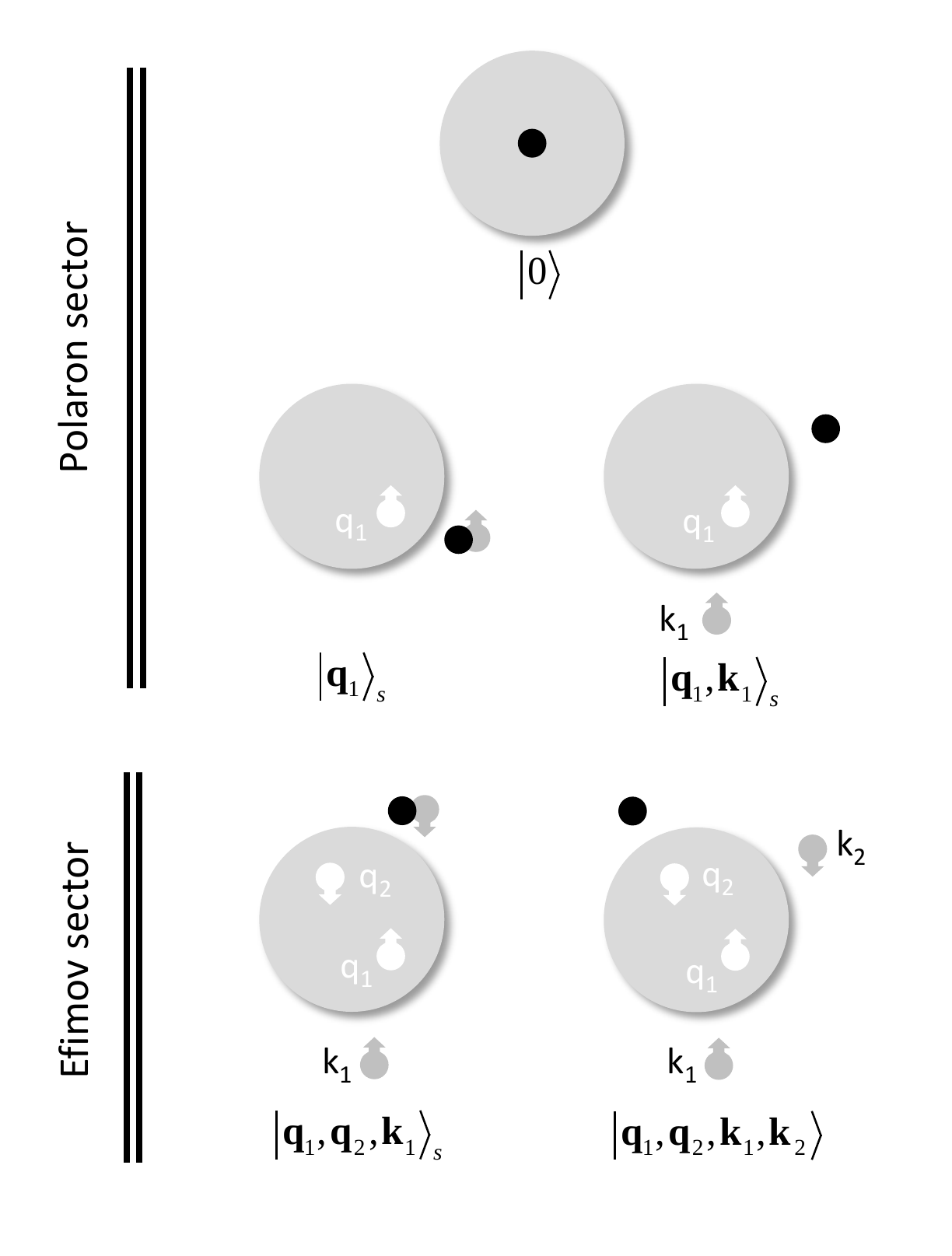}
    \caption{Structure of the variational Hilbert space. In the first two rows, the polaronic state is created by the impurity and one particle hole excitations, in the third row, a second particle hole excitation allows for the trimer to exist. }
    \label{fig:VariationalSpace}
\end{figure}

We search for the ground state energy within a variational space spanned by the states depicted in Fig. \ref{fig:VariationalSpace}. This space can be divided in two sectors. The polaron sector is spanned by state $|0\rangle$, which corresponds to the impurity sitting at the center of the two unperturbed Fermi seas, and single particle-hole states $|\bm q_1\rangle_s$ and $|\bm q_1,\bm k_1\rangle_s$ where a hole of spin $s$ and momentum $\bm q_1$ is accompanied by either a bound or  unbound impurity-fermion pair. The Efimov sector is characterized by states $|\bm q_1,\bm q_2,\bm k_1\rangle_s$ and $|\bm q_1,\bm q_2,\bm k_1,\bm k_2\rangle$ containing both one hole in each Fermi sea.

The general structure of a variational state is therefore
\be
\begin{split}
|\psi\rangle&=A|0\rangle + \sum_{\bm q_1,s}B_s(\bm q_1)|\bm q_1\rangle_s
+\sum_{\bm q_1,\bm k_1,s}C_s(\bm q_1,\bm k_1)|\bm q_1,\bm k_1\rangle_s\\&
 +\sum_{\bm q_1,\bm q_2,s}D_s(\bm q_1,\bm q_2,\bm k_1)|\bm q_1,\bm q_2,\bm k_1\rangle_s\\&+\sum_{\bm q_1,\bm q_2,\bm k_1,\bm k_2}E(\bm q_1,\bm q_2,\bm k_1,\bm k_2)|\bm q_1,\bm q_2,\bm k_1,\bm k_2\rangle.
\end{split}
\ee
with $q_i<k_F$ and $k_i>k_F$ and where $k_F$ is the Fermi wave-vector of the background fermions. For spin balanced Fermi seas, we can assume that the amplitudes do not depend on $s$ and within this subspace, we  explore two  families of variational states.

%\section{The polaron sector}
The polaronic sector corresponds to $D=E=0$. The corresponding Ansatz generalizes the approach succesfully used to describe the Fermi polaron problem, ie an impurity immersed in a spin-polarized Fermi sea \cite{chevy2006upa}. In particular this trial wavefunction recovers the exact perturbative expansion of the energy of the polaron up to second order in scattering length. In the following, we will assume that the impurity has the same mass as the fermions: $m_i=m$.

The minimization of the energy $W$  with respect to $A$, $B$ and $C$ in the polaronic sector can be reduced to a single scalar equation  $P_W=0$ with

\be
P_W=W-\frac{2}{\Omega}\sum_{q<k_F}\frac{1}{\Delta_{\bm q}(W)}
\label{Eq:polaron}
\ee
and
%\be
%\Delta_{\bm q}(W)=\frac{1}{\Lambda^2}(W-(E_0-q^2/4))-\frac{1}{\Omega}\sum_{k>k_F}\frac{1}{W-k^2+\bm k\cdot\bm q}\nonumber
%\label{Eq:W}
%\ee
\be
\begin{split}
\Delta_{\bm q}(W)=\frac{m}{4\pi\hbar^2}\bigg\{
a^{-1}-R_e(\lambda^2-\frac{q^2}{4})-\frac{2}{\pi}k_F\\
+\frac{4\pi}{\Omega}\sum_{k>k_F}\frac{1}{\lambda^2+k^2-\bm k\cdot\bm q}-\frac{1}{k^2}\bigg\}&
\label{Eq:W}
\end{split}
\ee
where $W = -\hbar^2\lambda^2/m $, is the ground state energy.

%Since in this sector the impurity is coupled to a single atom of the Fermi sea, the zero-range limit does not suffer from any singularity and in this regime the polaron energy is a universal function of the dimensionless parameter  $1/k_Fa$ only that we display in Fig. \ref{fig:polaron}.

In the perturbative regime $k_Fa\rightarrow 0^-$, the energy of the polaron can be expanded as
\be
W=\frac{8\pi\hbar^2a}{m}n_F\left(1+\frac{3}{2\pi}k_F a+...\right).
\ee
At this order, the variational result recovers the exact perturbative expansion and amounts to twice the interaction energy with a single spin component. Because of the non-linearity of Eq. (\ref{Eq:polaron}), this coincidence does not extend beyond that order. For instance, at the unitary limit $|a|=\infty$, we know that for a single component Fermi sea, the energy of the polaron is $W_{\rm FP}\simeq -0.606~E_F$ \cite{chevy2006upa,lobo2006nsp,prokof'ev08fpb}, where $E_F=\hbar^2 k_F^2/2m$ is the Fermi energy of the background fermions, while for a two component system, we find  $|W|\simeq 1.026 E_F<2 |W_{\rm FP}|$, meaning that, contrary to the perturbative expression, the interaction energy of the polaron with the two Fermi seas is not additive in the strong coupling regime.

We now consider the opposite limit $A=B=C=0$ corresponding to the formation of a ground state Efimov trimer above the Fermi surface.

As a reference, we first consider the energy of the trimer in the absence of a Fermi sea that is obtained as a solution of Skornyakov-Ter-Martirosyan's equation \cite{naidon2017efimov}
% \be
% \begin{split}
% \Bigg [\frac{1}{\Lambda^2}&(W+E_0+3k^2/4)\\
% -&\frac{1}{\Omega}\sum_{k'}\frac{1}{W-k^2-{k'}^2-\bm k\cdot\bm k'}\Bigg]D(k)\\
% +&\frac{1}{\Omega}\sum_{k'}\frac{D(k')}{W-k^2-{k'}^2-\bm k\cdot\bm k'}=0
% \end{split}
% \label{Eq:EfimovTrimer}
% \ee
%Modified by Ragheed (sign problem)
%\be
%\begin{split}
%\Bigg [\frac{1}{\Lambda^2}&(\lambda^2- \frac{1}{R_e a} + %\frac{2}{\pi R_e} k_F+3k^2/4)\\
%+&\frac{1}{\Omega}\sum_{k'}(\frac{1}{{k'}^2} - %\frac{1}{\lambda^2+k^2+{k'}^2+\bm k\cdot\bm k'})\Bigg]D(k)\\
%-&\frac{1}{\Omega}\sum_{k'}\frac{D(k')}{\lambda^2+k^2+{k'}^2%+\bm k\cdot\bm k'}=0\nonumber
%\end{split}
%\label{Eq:EfimovTrimer}
%\ee
\be
\begin{split}
\Bigg [\frac{1}{4\pi}&\bigg\{a^{-1}-R_e(\lambda^2+\frac{3}{4}k^2)
\bigg\}\\
+&\frac{1}{\Omega}\sum_{k'}( \frac{1}{\lambda^2+k^2+{k'}^2+\bm k\cdot\bm k'}-\frac{1}{{k'}^2})\Bigg]D(k)\\
+&\frac{1}{\Omega}\sum_{k'}\frac{D(k')}{\lambda^2+k^2+{k'}^2+\bm k\cdot\bm k'}=0
\end{split}
\label{Eq:EfimovTrimer}
\ee
%\textcolor{red}{XL: GLOBAL MINUS SIGN ?}
In this case, the only relevant dimensionless parameter is $R_e/a$ and we observe that the trimer merges with the atomic continuum for a scattering length $a_-$ such that $R_e/a_-=-2\times 10^{-4}$. It means that in our situation, where only the impurity-fermion interactions are resonant, the three-body bound states essentially  exist only in a regime where an impurity-fermion bound-state is also stable. This is to be contrasted with the more traditional three-boson problem for which all three interactions are resonant and Efimov trimers are stable deep in the domain where two-body bound states are unstable  (in this case we have indeed $R_e/a_-\simeq -0.1$).

\begin{figure}
    \centering
    \includegraphics[width=\columnwidth]{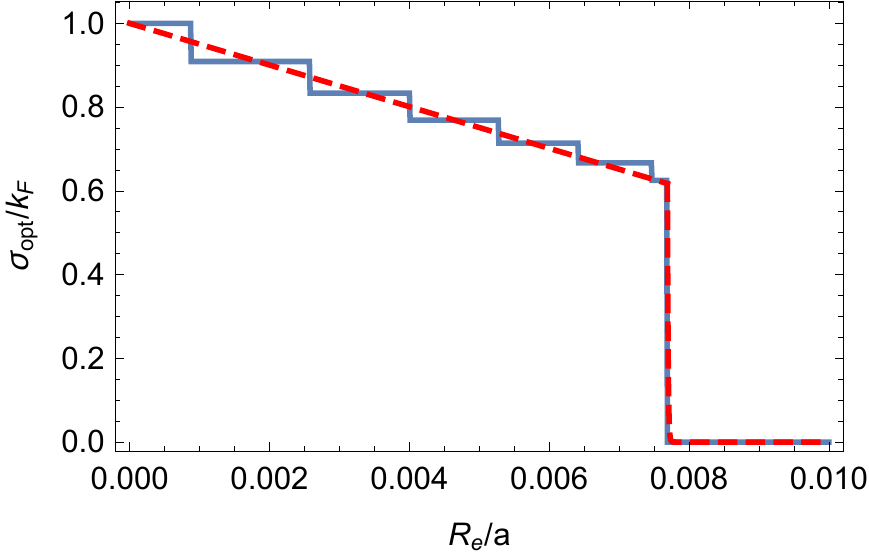}

    \includegraphics[width=\columnwidth]{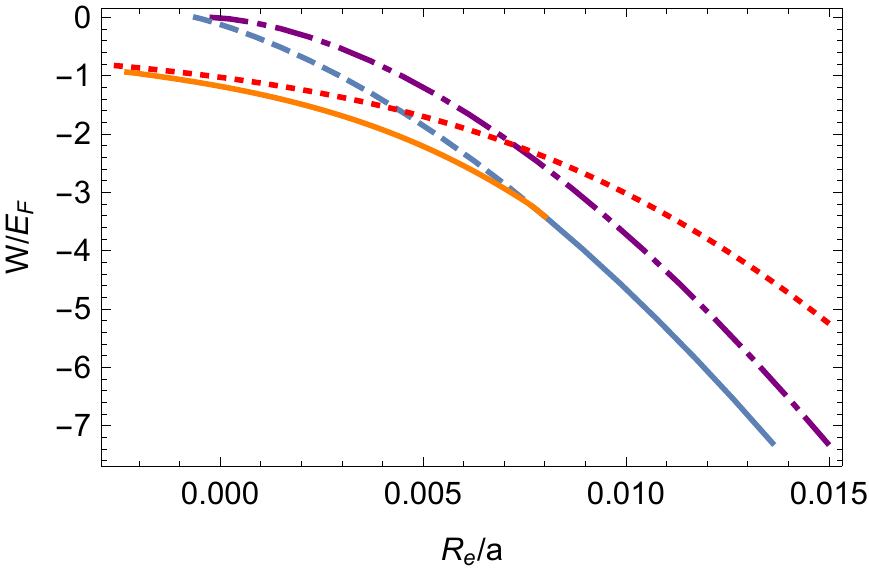}
    \caption{Variational ground state in the polaron-trimeron space for $k_F R_e=10^{-2}$. Upper panel: optimal value of the width $\sigma$ of the hole-pair wavefunction (blue). $\sigma$ was varied over a finite set of values between between 0 and 1 (see text). For $R_e/a\lesssim 0.0077$, each step corresponds to a jump from one value of sigma to the next and is therefore an artefact of the discretization of $\sigma$. The red-dashed line corresponds to a smooth interpolation. For $R_e/a \simeq 0.0077$, We observe a jump of the value of $\sigma$ which straddles several consecutive values of $\sigma$ and thus marks a discontinuity  between a Cooper-like trimer and a polaron-trimeron crossover state. Lower panel: energy of the variational state.    Cooper-like trimer ($\sigma=0$, blue), compared to the energy of an Efimov trimer in vacuum (purple, dash-dotted). The ground state associated with the optimal value of $\sigma$ is the solid line. The orange section of the line corresponds to finite width hole pair wave functions $F$ while the blue section corresponds  to  Cooper-like trimers ($\sigma=0$).}
    \label{fig:VariationalEnergy}
\end{figure}

We  consider next the effect of the Fermi sea on the energy of the trimer. In a first approach we simply assume that its role is to prevent the fermions above the Fermi surface to occupy states below $k_F$, in a manner very similar to the celebrated Cooper pairing problem for pairs of fermions in superconductors. These “Cooper-like" trimer states correspond to locating hole momenta $\bm q_{1,2}$ on the Fermi surface, and having $\bm q_1+\bm q_2=0$ to cancel the center of mass momemtum of the trimer. The energy of the trimer state is then solution of:

% \be
% \begin{split}
% \Bigg [\frac{1}{\Lambda^2}&(W-2E_F+E_0+3k^2/4)\\
% -&\frac{1}{\Omega}\sum_{k'>k_F}\frac{1}{W-2E_F-k^2-{k'}^2-\bm k\cdot\bm k'}\Bigg]D(k)\\
% +&\frac{1}{\Omega}\sum_{k'>k_F}\frac{D(k')}{W-2E_F-k^2-{k'}^2-\bm k\cdot\bm k'}=0
% \end{split}
% \label{Eq:CooperTrimer}
% \ee
%Modified by Ragheed
%\be
%\begin{split}
%\Bigg [\frac{1}{\Lambda^2}&(\lambda^2 - k_F^2- \frac{1}{R_e a} + \frac{2}{\pi R_e} k_F+3k^2/4)\\
%+&\frac{1}{\Omega}\sum_{k'}(\frac{1}{{k'}^2} - %\frac{1}{\lambda^2 - k_F^2+k^2+{k'}^2+\bm k\cdot\bm %k'})\Bigg]D(k)\\
%-&\frac{1}{\Omega}\sum_{k'>k_F}\frac{D(k')}{\lambda^2 - %k_F^2+k^2+{k'}^2+\bm k\cdot\bm k'}=0
%\nonumber
%\end{split}
%\label{Eq:CooperTrimer}
%\ee
\be
\begin{split}
\Bigg [
\frac{1}{4\pi}&\bigg\{a^{-1}-\frac{2}{\pi}k_F-R_e(\lambda^2- k_F^2+\frac{3}{4}k^2)\bigg\}
\\
+&\frac{1}{\Omega}\sum_{k'>k_F}(\frac{1}{\lambda^2 - k_F^2+k^2+{k'}^2+\bm k\cdot\bm k'}-\frac{1}{{k'}^2})\Bigg]D(k)\\
+&\frac{1}{\Omega}\sum_{k'>k_F}\frac{D(k')}{\lambda^2 - k_F^2+k^2+{k'}^2+\bm k\cdot\bm k'}=0
\end{split}
\label{Eq:CooperTrimer}
\ee
This equation is very similar to Eq. (\ref{Eq:EfimovTrimer}), the main difference stemming from the sums over momenta that are now restricted to $k>k_F$ and the shift of the energy associated with the chemical potential of the two fermions that were removed from the Fermi seas  to create the trimer. The corresponding ground state energy is plotted in Fig. \ref{fig:VariationalEnergy} for an experimentally relevant value $k_F R_e\simeq 10^{-2}$. We observe that like for traditional Cooper pairing the presence of the Fermi sea stabilizes the trimer. %In particular, while in vacuum Efimov trimers do not exist beyond a critical scattering length $a_-$ (see above), the presence of the Fermi Sea stabilizes the trimer for arbitrarily weak attraction, albeit with an exponentially small binding energy.

We can generalize this result by considering trimer amplitudes $D_s$ and $E$ of the form

\begin{eqnarray}
D_s(\bm q_1,\bm q_2,\bm k_1)&=&F(\bm q_1,\bm q_2)\tilde D(\bm k_1)\\
E(\bm q_1,\bm q_2,\bm k_1,\bm k_2)&=&F(\bm q_1,\bm q_2)\tilde E(\bm k_1,\bm k_2),
\end{eqnarray}
where the Cooper-like trimer corresponds to $F(\bm q_1,\bm q_2)=\delta_{\bm q_1,-\bm q_2}\widetilde F(q_1)$, where $\widetilde F$ is peaked near the Fermi surface.  We choose the following normalization for the function $F(\bm q_1,\bm q_2)$: $\sum_{q_1,q_2} |F(\bm q_1,\bm q_2)|^2 = N_F^2$, where $N_F$ is the total number of fermions per spin state.

Once again, we can eliminate $E$ and we see that at fixed $F$, $\tilde D$ is solution of a Skornyakov-Ter-Martirosyan like equation :
%Aussi modifié par Ragheed
%\be
%\begin{split}
%\Bigg [\frac{1}{\Lambda^2}&(\lambda^2- \frac{1}{R_e a} + %\frac{2}{\pi R_e} k_F+3k^2/4-\langle (\bm q_1-\bm %q_2)^2\rangle/4)\\
%-&\frac{1}{\Omega}\sum_{k'>k_F}\frac{1}{\lambda^2+\langle\bm %q_1\cdot\bm q_2\rangle+k^2+{k'}^2+\bm k\cdot\bm %k'}\Bigg]D(k)\\
%-&\frac{1}{\Omega}\sum_{k'>k_F}\frac{D(k')}{\lambda^2+\langl%e\bm q_1\cdot\bm q_2\rangle+k^2+{k'}^2+\bm k\cdot\bm %k'}=0\nonumber
%\end{split}
%\label{Eq:trimeronF}
%\ee
\be
\begin{split}
\Bigg [
\frac{1}{4\pi}&\bigg\{a^{-1}-\frac{2}{\pi}k_F-R_e(\lambda^2+\frac{3}{4}k^2-\dfrac{\langle (\bm q_1-\bm q_2)^2\rangle}{4})+\bigg\}
\\
+&\frac{1}{\Omega}\sum_{k'>k_F}(\frac{1}{\lambda^2+\langle\bm q_1\cdot\bm q_2\rangle+k^2+{k'}^2+\bm k\cdot\bm k'}-\frac{1}{{k'}^2})\Bigg]\tilde{D}(k)\\
+&\frac{1}{\Omega}\sum_{k'>k_F}\frac{\tilde{D}(k')}{\lambda^2+\langle\bm q_1\cdot\bm q_2\rangle+k^2+{k'}^2+\bm k\cdot\bm k'}=0
\end{split}
\label{Eq:trimeronF}
\ee
with $\langle f(\bm q_1,\bm q_2)\rangle=\sum_{\bm q_i}|F(\bm q_1,\bm q_2)|^2 f(\bm q_1,\bm q_2)/N_F^2$, and where we assumed that the distribution $|F|^2$ is an even function of $\bm q_1$ and $\bm q_2$. Comparing Eq. (\ref{Eq:CooperTrimer}) and (\ref{Eq:trimeronF}) we see that their respective energies are simply translated one with respect to the other since we have:
% \be
% \begin{split}
% W_F(R_e/a)=&W_C(R_e/a+R_e^2\langle(\bm q_1+\bm q_2)^2\rangle/4)\\
% +&\frac{\hbar^2}{m}\left(k_F^2+\langle\bm q_1\cdot\bm q_2\rangle\right).
% \end{split}
% \ee
%Modified by ragheed
\be
\begin{split}
W_F(R_e/a)=&W_C(R_e/a+R_e^2\langle(\bm q_1+\bm q_2)^2\rangle/4)\\
-&\frac{\hbar^2}{m}\left(k_F^2+\langle\bm q_1\cdot\bm q_2\rangle\right).
\end{split}
\ee
This mapping corresponds to a translation of both the argument and the value of $W_c$ and in practice we observe that the latter dominates. Since $q_1$ and $q_2$ are bounded by the Fermi wavevector $k_F$, we see that $k_F^2+\langle\bm q_1\cdot\bm q_2\rangle$ is always positive and the Cooper-like Ansatz is always the optimal choice %(we illustrate this property in Fig. \ref{fig:VariationalEnergy} for a constant $F$).
We now study the hybridization of the polaronic and Efimov sectors by minimizing the energy with respect to all five amplitudes $A, B, C, D$ and $E$.  From the previous analysis, we would expect that the optimal choice would be to mix the polaron wavefunction with the Cooper-like trimer. However, as we will show below, these two sectors are not coupled at the thermodynamic limit. Indeed, the normalization of the state of a Cooper-like trimer requires that
\be
\begin{split}
|A|^2&+2\sum_{\bm q_1}|B(\bm q_1)|^2+2\sum_{\bm q_1,\bm k_1}|C(\bm q_1,\bm k_1)|^2\\
+&2\sum_{\bm q_1,\bm k_1}|D(\bm q_1,-\bm q_1,\bm k_1)|^2\\
+&\sum_{\bm q_1,\bm k_1,\bm k_2}|E(\bm q_1,-\bm q_1,\bm k_1,\bm k_2)|^2=1
\end{split}
\ee
For large quantization volumes, the sums are turned into integrals and to recover results that do no depend on $\Omega$, we see that $A$ should not depend on $\Omega$ and $B$, $C$, $D$ and $E$ should respectively scale like
\begin{eqnarray}
&B=\dfrac{b(\bm q_1)}{\sqrt{\Omega}}, \, D(\bm q_1,-\bm q_1,\bm k_1)=\dfrac{d(\bm q_1,\bm k_1)}{\Omega}\\
&C=\dfrac{c(\bm q_1,\bm k_1)}{\Omega}, \, E(\bm q_1,-\bm q_1,\bm k_1,\bm k_2)=\dfrac{e(\bm q_1,\bm k_1,\bm k_2)}{\Omega^{3/2}}
\nonumber
\end{eqnarray}
where $b$, $c$, $d$ and $e$ do not depend on the size of the system. Under this assumption, the interaction term of the Hamiltonian can be recast as
\be
\begin{split}
\langle\widehat H_{\rm int}\rangle=&\Lambda\Huge[\int\frac{d^3\bm q_1}{(2\pi)^3}A^*b(\bm q_1)\\
+&\int \frac{d^3\bm q_1 d^3\bm k_1}{(2\pi)^6}b(\bm q_1)^*c(\bm q_1,\bm k_1)\\
+&\int \frac{d^3\bm q_1 d^3\bm k_1}{(2\pi)^6\sqrt{\Omega}}c(\bm q_1,\bm k_1)^*d(\bm q_1,\bm k_1)\\
+&\int \frac{d^3\bm q_1 d^3\bm k_1 d^3\bm k_2}{(2\pi)^9}d(\bm q_1,\bm k_1)^*e(\bm q_1,\bm k_1,\bm k_2)\Huge]\\
+&{\rm c.c.}
\end{split}
\ee
In this expression, we see that the energy does not depend on the quantization volume, except for the term coupling the amplitudes $c$ and $d$ which vanishes as $1/\sqrt{\Omega}$ for diverging $\Omega$ thus showing that in this  limit, the polaron and Cooper-like trimer sectors are  decoupled.

To explore a possible polaron-trimeron crossover we therefore need to relax the constraint on the vanishing center of mass momentum characterizing the Cooper-like trimer state. For this purpose we consider a trial wavefunction $F(\bm q_1,\bm q_2)=F_0e^{-\bm q_1\cdot\bm q_2/2\sigma^2}$, where $F_0$ is a normalization constant. Just like for the Cooper-like trimer, this amplitude is maximum when $\bm q_1+\bm q_2=0$ and when both momenta are on the Fermi surface. The parameter $\sigma$ allows us to tune continuously the width of the hole wave-function between a uniform distribution and the Cooper-like trimer configuration. The Cooper-like trimer corresponds to $\sigma=0$ while the opposite limit ($\sigma=\infty$) corresponds to a uniform distribution $F$.

The minimization of the energy with respect to the amplitudes $A,B,...,E$ yields the following  set of coupled equations on $\tilde{A}=A/(\sqrt{\Omega}N_F\Lambda)$ and $\tilde{D}$ generalizing Eq. (\ref{Eq:polaron}) and (\ref{Eq:trimeronF})
\begin{eqnarray}
P_W \tilde{A}&=&\frac{2}{\Omega}\sum_{{\bf k}} h(k)\tilde{D}(k)\label{Eq:coupledEq1}\\
{\cal T}[\tilde{D}](k)&=& h(k)\tilde{A} + f(k) \tilde{D}(k) + \frac{1}{\Omega}\sum_{{\bf k}'} g({\bf k},{\bf k}')\tilde{D}(k'), \nonumber\\
\label{Eq:coupledEq2}
\raisetag{10pt}
\end{eqnarray}
where ${\cal T}$ is the operator from Eq. (\ref{Eq:trimeronF}) (times $m/\hbar^2$), and the coupling functions $h, f , g$ are the following:
% \be
% \begin{split}
% &h(k) = \frac{1}{\Omega}\sum_{{\bf q}} \frac{\beta(q)}{(\lambda^2 + k^2 + \bm q\cdot\bm k)\Delta_{\bm q}},\\
% &f(k) =-\frac{1}{\Omega} \sum_{{\bf q}} \frac{\beta(q)^2}{\lambda^2 + k^2 - \bm q\cdot\bm k},\\
% &g({\bf k},{\bf k}') = -\frac{1}{\Omega}\sum_{{\bf q}} \frac{\beta(q)^2}{(\lambda^2 + k^2 - \bm q\cdot\bm k)(\lambda^2 + k'^2 - \bm q\cdot\bm k')\Delta_{\bm q}},\\
% % &\Delta(q) = \frac{1}{\Lambda^2}(\lambda^2 -\frac{1}{R_e a} - \frac{q^2}{4}) + \sum_k (\frac{1}{k^2} -\frac{1}{(\lambda^2 + k^2 - \bm q\cdot\bm k)}),\\
% &\beta(q) = \frac{1}{N_F}\sum_{{\bf q}'} F({\bf q},{\bf q}')
% \nonumber
% \end{split}
% \ee
\be
\begin{split}
&h(k) = -\frac{m}{\hbar^2\Omega}\sum_{{\bf q}} \frac{\beta(q)}{(\lambda^2 + k^2 - \bm q\cdot\bm k)\Delta_{\bm q}},\\
&f(k) =-\frac{m}{\hbar^2\Omega} \sum_{{\bf q}} \frac{\beta(q)^2}{\lambda^2 + k^2 - \bm q\cdot\bm k},\\
&g({\bf k},{\bf k}') = \frac{m^2}{\hbar^4\Omega}\sum_{{\bf q}} \frac{\beta(q)^2}{(\lambda^2 + k^2 - \bm q\cdot\bm k)(\lambda^2 + k'^2 - \bm q\cdot\bm k')\Delta_{\bm q}},\\
% &\Delta(q) = \frac{1}{\Lambda^2}(\lambda^2 -\frac{1}{R_e a} - \frac{q^2}{4}) + \sum_k (\frac{1}{k^2} -\frac{1}{(\lambda^2 + k^2 - \bm q\cdot\bm k)}),\\
&\beta(q) = \frac{1}{N_F}\sum_{{\bf q}'} F({\bf q},{\bf q}')
\label{Eq:auxfuncs}
\end{split}
\ee
and where $\Delta_{\bm q}$ is defined in Eq. (\ref{Eq:W}).

The results of the minimization are displayed in Fig. \ref{fig:VariationalEnergy} \cite{SuppMat}. For each value of $R_e/a$ we  solve Eq. (\ref{Eq:coupledEq1}) and (\ref{Eq:coupledEq2}) for a fixed set of values of $\sigma$ ($\sigma\in\{0, 0.1, 0.125, 0.2, 0.25, 0.4, 0.5, 0.625, 0.67, 0.71,
0.77, 0.83,\\ 0.91, 1\}$, $\sigma =0$ corresponding to the Cooper-like trimer).  For each value of $R_e/a$ we search for the optimal value of $\sigma_{\rm opt}$ that minimizes the energy of the impurity. The corresponding values of $\sigma_{\rm opt}$ are displayed in Fig.  (\ref{fig:VariationalEnergy}). For $R_e/a\lesssim 7.7 \times 10^{-3}$, we observe that the optimum value decreases smoothly (in this regime the steps are just due to the discrete values of $\sigma$) and drops to $\sigma_{\rm opt} =0$ (corresponding to the Cooper-like trimer) at $R_e/a\simeq 7.7\times 10^{-3}$ that suggests a sharp transition to the Cooper-like trimer state.
The variational ground state energy  corresponding to $\sigma=\sigma_{\rm opt}$ is displayed in the lower panel of Fig. \ref{fig:VariationalEnergy}, as well as the energy of the polaron state, and of the Cooper-like trimer. On this graph, we clearly see that for weak attractive interactions ($R_e/a\rightarrow-\infty$), the variational states converges to the polaron energy and that the ground state abruptly jumps to the Cooper-like trimer state in the vicinity of $R_e/a\simeq 7.7\times 10^{-3}$. Note that this critical value depends on $k_F R_e$ and should converge to $R_e/a_-$ for vanishing fermionic density.

%\section{Conclusion}

From our variational approach, we conclude that the ground state of an impurity immersed in a non-interacting mixture of spin 1/2 fermions undergoes a first-order transition between a polaronic and a trimer state. This is different from the case of superfluid background where the presence of Cooper pairs allows for a crossover between these two states. We attribute this difference to the absence of coupling at the thermodynamic limit between the vector space spanned by the polaron and that of the trimer state. In the strongly attractive limit, the energy of the Cooper-like trimer is lowered thanks to the exact cancellation of its center of mass motion. A similar situation seems to occur in the Fermi polaron for the transition between the polaron and dimer states \cite{lan2014single} where the dimer state was described in a variational subspace containing two particle-hole pairs located on the Fermi surface, a question that is still open \cite{edwards2013smooth,ness2020observation,cui2020fermi}. More generally our approach can help understanding the transition between few body states in many-body environment, such as the recently reported trimer/dimer avoided crossing for the Efimov ground state \cite{yudkin2020efimov}. Finally, a natural extension of this work would be to study the existence of such a transition in the case of the {\em normal} state of an interacting fermionic background to understand the relative role of pairing and superfluidity in this problem.

\acknowledgments
The authors thank F. Werner, K. van Houcke, S. Giorgini, C. Lobo, L. Khaykovich and C. Salomon  for helpful discussions. F. Chevy and R. Alhyder acknowledge support from ERC (advanced grant CritiSup2) and Fondation Simone et Cino del Duca.

\bibliographystyle{unsrt}
\bibliography{bibliographie}
\newpage
\section{Numerical calculation}
The energy spectrum of Eq. (\ref{Eq:coupledEq2}) can be obtained numerically. In the following we put $\hbar$ and $m$ to unity for simplicity. We use the following parameters:
\begin{align}
\begin{split}
x= \lambda R_e&, \quad y = k_F/\lambda\\
 k = \lambda\sinh(\xi)&,\quad \phi(\xi) = \lambda\sinh(\xi)D(\lambda\sinh(\xi))
\label{Eq:24}
\end{split}
\end{align}
With this we can write Eq. (\ref{Eq:coupledEq2}) as follows:
 \begin{align}
\begin{split}
\mathcal{M}(x,y,a) \tilde{\phi} = 0
 \label{eq:25}\end{split}
 \end{align}
Where $\mathcal{M}$ is a non-symmetric square matrix.  Its elements have the following form:
 \begin{align}
\begin{split}
&\mathcal{M}_i(\xi,\xi',x,y,a) = \delta(\xi,\xi')\mathcal{M}_D(\xi,x,y,a)\\&+ \epsilon \int_0^{2\pi}d\phi\int_0^\pi d\theta \sin(\theta) \cosh(\xi')\sinh^2(\xi')\mathcal{M}_N(\xi,\xi',x,y,a)
\raisetag{42pt}
 \label{eq:26}\end{split}
 \end{align}
 Where $\epsilon$ is a discretization step we choose in order to have a convergent value of the unknown variable. We note that $\epsilon = \xi'_{max}/N_p$ where $\xi'$ is the cutoff of the integral over $\xi'$ and $N_p$ is the number of points per row.  Note also that $\xi, \xi' > \text{asinh}(y)$. We define the two functions $\mathcal{M}_D$ and $\mathcal{M}_N$ as follows:
  \begin{align}
\begin{split}
&\mathcal{M}_N(\xi,\xi',x,y,a) = \frac{1}{ \lambda^2I(\xi,\xi')} + \frac{2h(\xi) h(\xi') }{P_W} + g(\xi,\xi')
\raisetag{7pt}
 \label{eq:27}\end{split}
 \end{align}
 and
  \begin{align}
\begin{split}
&\mathcal{M}_D(\xi',x,y,a) = f(\xi)- \frac{\lambda}{4\pi}\bigg\{\Big(1+\frac{3}{4}\sinh(\xi)^2\\&-\langle (\bm q_1-\bm q_2)^2\rangle/(4\lambda^2))\Big)\lambda R_e- \frac{1}{\lambda a}+\frac{2k_F}{\pi\lambda}\bigg\}
\\&-\frac{1}{\lambda^3\Omega}\sum_{\xi'>\text{asinh}(y)}\frac{1}{I(\xi,\xi')}\Bigg]
 \label{eq:28}\end{split}
 \end{align}
Where:
\begin{align}
\begin{split}
&I(\xi,\xi')=1+\langle\bm q_1\cdot\bm q_2\rangle/\lambda^2+\sinh^2(\xi)\\&+\sinh^2(\xi')+\sinh(\xi)\sinh(\xi')\cos(\theta)
 \label{eq:29}\end{split}
 \end{align}
 and $g$ and $h$ are functions defined in Eq. \eqref{Eq:auxfuncs} , after the proper variable change. By fixing the value of $k_FR_e$ and for each value of $x = x_c$, we search the smallest value of $1/\lambda a$ which verifies the equation:
  \begin{align}
\begin{split}
\det [\mathcal{M}(x_c,y_c,a)] = 0
 \label{eq:30}\end{split}
 \end{align}
 Note that the results in the Fig. \ref{fig:VariationalEnergy} are obtained using a total number of points $N_p = 350$ and $\xi'_{max} = 10$.

\end{document}